\documentclass[]{aa}  

\usepackage{graphicx}
\usepackage{txfonts}
\usepackage{hyperref}
\hypersetup{
    colorlinks=true,
    linkcolor=blue,
    citecolor=blue,
    filecolor=magenta,      
    urlcolor=cyan,
    }

\begin{document}

\title{Unveiling stellar spin: Determining inclination angles in Be stars}

\subtitle{}
   \author{D. Turis-Gallo  \inst{1},
   M. Cur\'e \inst{1},  
   R. S. Levenhagen \inst{2},
   C. Arcos \inst{1}, 
   I. Araya  \inst{3},
   \and 
   A. Christen  \inst{4}
                    }
\institute{Instituto de F\'isica y Astronom\'ia, Facultad de Ciencias, Universidad de Valpara\'iso,  Av. Gran Bretaña 1111, Valpara\'iso, Chile.\\
\email{daniela.turis@postgrado.uv.cl} \and
Departamento de Física, Universidade Federal de São Paulo, Rua Prof. Artur Riedel, 275, 09972-270, Diadema, SP, Brazil. \and
Centro Multidisciplinario de F\'isica, Vicerrector\'ia de Investigaci\'on, Universidad Mayor, 8580745 Santiago, Chile \and
Instituto de Estad\'istica, Facultad de Ciencias, Universidad de Valpara\'iso,  Av. Gran Bretaña 1111, Valpara\'iso, Chile.
}
\date{}
 
\abstract
 {The physical properties of stellar atmospheres in rapidly rotating massive stars, such as Be stars, are critical to understanding their evolution and their role as progenitors of supernovae. 
 These stars, which often have near-critical rotation, exhibit equatorial stretching and gravity darkening, which significantly complicates the determination of parameters such as the inclination angle. 
  Be stars, characterized by their extreme rotational velocities, serve as excellent candidates for exploring these phenomena. However, fundamental quantities such as polar and equatorial radii and inclination angles are typically derived from interferometry, which applies only to a limited number of stars.}
 {This study aims to enhance the determination of inclination angles for Be stars using the \texttt{ZPEKTR} spectral synthesis code. By incorporating advanced models of gravity 
 darkening and stellar deformation, we evaluated the effectiveness of this method with a sample of ten Be stars from the BeSOS database, comparing results with established interferometric data.}
{We used the \texttt{ZPEKTR} code to model the effects of stellar oblateness and gravity darkening on spectral lines, focusing on the HeI 4471 \AA\ line. We applied a $\chi^{2}$-test minimization approach to identify the best-fitting models, and we evaluated the inclination angles derived against interferometric measurements.  
}
   {Our analysis reveals a robust linear correlation (slope: $0.952 \pm 0.033$; $R^2 = 0.989$)  between the inclination angles derived from \texttt{ZPEKTR} and using interferometric techniques, 
 which demonstrates an excellent agreement. The \texttt{ZPEKTR} code effectively models high rotational velocity effects, providing precise stellar parameter determinations.}   
{The \texttt{ZPEKTR} code is a powerful tool for estimating inclination angles in Be stars. The results underscore the potential of advanced spectroscopic techniques 
to yield inclination measurements comparable to interferometry, which offers a pathway to studying distant massive stars for which interferometric observations are not feasible.}
  
   
\titlerunning{Unveiling the Stellar Spin}\authorrunning{Turis-Gallo et al.}
\maketitle
\section{Introduction}

Classical Be stars are fast-rotating B-type non-supergiant stars characterized by prominent emission features in Balmer lines, often accompanied by emissions in optically thin metal lines. These features are commonly interpreted as being the result of radiative re-emission processes occurring within a disc-like gaseous circumstellar environment in Keplerian rotation \citep{Rivinius2013}. While the precise mechanisms responsible for forming such circumstellar structures remain an open question, there is broad consensus that the stellar mass loss is driven by a coupled mechanism involving fast stellar rotation and non-radial pulsations \citep{Rivinius2013b,Labadie2021}.

Be stars rotate close to their critical rotation \citep{Meilland2012, Zorec2016}, which significantly impacts their evolutionary paths in the Hertzsprung-Russell diagram \citep{Ekstrom2008, Granada2013} and their observable properties, such as gravity-darkening effects {\citep{Fremat2005,Espinosa-Lara}}, measured parameters \citep{Abdul-Masih2023}, and polarization \citep{Bailey2020,Bailey2024}. 
A consequence of rapid rotation is the very broad helium and metal line profiles observed in the spectra of these stars. Moreover, in spectroscopic observations, we measure the projected rotational velocity $(v\,\sin{i)}$ rather than the true equatorial velocity ($v)$ because only the component of rotational velocity projected along the line of sight contributes to Doppler broadening, making the inclination angle ($i$) a key but often unknown factor in determining the actual rotational speed of the star. Therefore, most Be stars are observed with high $v\,\sin{i}$ values, as the probability of detecting them at a given inclination angle is determined by $dP = \sin{i} \, di$, which favours higher inclination angles relative to the observer's line of sight.

Determining inclination angles is fundamental for characterizing rapidly rotating stars, providing critical insights into their geometry, rotation, and circumstellar environments. Inclination angles allow for the derivation of key stellar properties, including the equatorial rotational velocity and the orientation of the star-disc system relative to the observer. Interferometric techniques are considered the gold standard for determining inclination angles, as they directly resolve stellar discs and provide precise geometric constraints \citep{Domiciano2003,Meilland2008,Kanaan2008,Meilland2011,Delaa2011,Cochetti2019,Klement2022}. However, such methods are observationally expensive and limited to nearby bright stars. 

Apart from direct measurements of stellar inclination angles by interferometry, there are at least three indirect approaches to their estimation. The longstanding direct inspection of Balmer lines in emission, in the case of Be stars, is very traditional and assumes the circumstellar envelope is thin and orbiting along the stellar equatorial region \citep{Arcos2017,Sigut2020,Marr2022,Sigut2023,Rubio2023,Lailey2024}. Another traditional method for inferring the positions of rotation axes is based on asteroseismological studies of stellar oscillation modes \citep{Gehan2021A&A...645A.124G}. A more recent method is inspired by the observations of the tilting of line profiles \citep{Lesage2014A&A...563A..86L}. On one hand, our method differs from the methods mentioned above because it relies on fitting observed line profiles with gravity-darkened models. On the other hand, it is similar to the work by \cite{Fremat2005} but with an updated gravity-darkening model \citep{Espinosa-Lara}. This spectroscopic approach is more widely applicable and can be used to study larger samples. Despite their potential, spectroscopic methods are subject to uncertainties introduced by rotational effects, and if they are not considered, errors can reach up to 40$\%$ of the $v\sin{i}$ measured \citep{Stoeckley}. 

In this context, the \texttt{ZPEKTR} code \citep{Levenhagen2014, Levenhagen2024A&A...685A..57L} incorporates advanced modelling of gravity darkening, limb darkening (LD), and geometric deformation due to rotation. Gravity darkening, initially described by \citet{VonZeipel} and later refined by \citet{Espinosa-Lara}, captures the latitudinal dependence of temperature and flux caused by the oblate shape of a rapidly rotating star. Limb-darkening models the variation in brightness across the stellar disc due to the angle of emission relative to the observer. Additionally, the geometric deformation modelled in \texttt{ZPEKTR} accounts for the departure of the stellar surface from spherical symmetry, a critical consideration for near-critical-rotation stars.

In this study, we utilized \texttt{ZPEKTR} to model the HeI 4471 \AA\ spectral line. This analysis was applied to a sample of Be stars from the Be Stars Observation Survey (BeSOS\footnote{\url{https://besos.ifa.uv.cl}}) catalogue  \citep{Arcos2018}, a high-resolution spectral database of massive stars. By comparing inclination angles derived from \texttt{ZPEKTR} with those obtained through interferometric methods \citep{Cochetti2019}, we aimed to evaluate the accuracy and reliability of spectroscopic approaches in determining the inclination angles of rapidly rotating stars. This work provides critical insights into the strengths and limitations of spectroscopic methods, and contributes to a broader understanding of the rotational dynamics and evolutionary pathways of massive stars.


\section{Data}

To assess the accuracy of the inclination angle measurements derived through a spectroscopic method using the \texttt{ZPEKTR} code, we selected ten Be stars with known inclination angles, determined interferometrically in the work of \cite{Cochetti2019}. Interferometric measurements provide a highly accurate benchmark for evaluating our spectroscopic approach because they directly resolve stellar discs and provide precise geometric constraints on stellar parameters, including inclination angles \citep[see e.g.][]{Sigut2020}.

The ten Be stars analysed in this study were selected based on the availability of high-resolution spectroscopic data in BeSOS, which primarily catalogues classical Be stars \citep{Arcos2018}. Observations were conducted between 2012 and 2015 using the Pontificia Universidad Católica High Echelle Resolution Optical Spectrograph (PUCHEROS), mounted on a 50 cm telescope at the Santa Martina Observatory. PUCHEROS, developed at the Center of AstroEngineering at Pontificia Universidad Católica, provides a spectral resolution of approximately 17000, with a limiting magnitude of $V = 9$ and a signal-to-noise ratio of 20 \citep{vanzi2012}.


\section{Method}

This section describes the ZPEKTR code, which we used to model the HeI 4471 \AA\ spectral line and evaluate the best-fitting models. ZPEKTR incorporates the Espinosa-Lara formalism and considers the impacts of fast star rotation. The code outputs essential stellar parameters, including radius, surface gravity, rotational velocity, and inclination angle. To identify the best-fitting models, we performed a $\chi^{2}$ test that compared synthetic spectra generated by ZPEKTR to observed HeI 4471 \AA\ line profiles. 

\subsection{\texttt{ZPEKTR} code}

The \texttt{ZPEKTR} code \citep{Levenhagen2014,Levenhagen2024A&A...685A..57L} was employed for spectral synthesis; we used it to model the effects of rapid stellar rotation on spectral lines. Assuming rigid rotation, the code incorporates the formalism of \cite{Espinosa-Lara}, which introduces a novel relationship between effective temperature and surface gravity that varies as a function of the latitudinal angle.

In the classical von Zeipel formulation, the flux divergence within the stellar core is assumed to be proportional to the nuclear energy generation rate, while outside the stellar core, the flux divergence is expected to be zero \citep{VonZeipel}. However, in the presence of rotation, the flux divergence in the outer layers deviates from this assumption, alternating between negative and positive values. These variations induce instabilities that drive the formation of meridional currents. On the other hand, the formalism developed by \cite{Espinosa-Lara} resolves the flux divergence issues present in the classical von Zeipel formulation, whose relationship is defined as follows \citep{Espinosa-Lara}:

\begin{equation} T_{\rm{eff}} = \left(\frac{F}{\sigma}\right)^{1/4} = \left( \frac{L}{4 \pi \sigma G M} \right)^{1/4} \sqrt{\frac{\tan{\vartheta}}{\tan{\theta}}} \,\, g_{\rm{eff}}^{1/4}, \label{eq:GD} \end{equation}
where $\sigma$ is the Stefan-Boltzmann constant, $G$ is the gravitational constant, $L$ and $M$ are the luminosity and mass of the star, respectively, $\theta$ is the co-latitudinal angle, and $\vartheta$ is defined as

\begin{equation}
     \cos{\vartheta} + \ln{\frac{\vartheta}{2}} = \frac{1}{3} \omega^{2} \tilde{r}^{3} \cos{\theta}^{3} + \cos{\theta} + \ln{\tan{\frac{\theta}{2}}} ,
\end{equation}
with $\tilde{r} = r / R_{eq}$, where $R_{eq}$ is the equatorial radius, and $\omega$ is the angular rotation rate  ($\Omega / \Omega_{c}$), defined as the ratio of the angular rotational velocity, $\Omega$, to the critical angular rotational velocity, $\Omega_{c}$. This approach resolves the flux divergence issues of the classical von Zeipel model by assuming that the radiative flux is always antiparallel to the effective gravity across all stellar layers. 

The \texttt{ZPEKTR} code also accounts for the geometrical deformations induced by rapid rotation, which result in a non-spherical shape for the star. Specifically, the code incorporates the latitudinal variation in the stellar radius as a function of rotational velocity, thereby improving the representation of the geometry of the stars in the gravitational potential calculation, which, at any colatitude of a rotating star, is

 \begin{equation}
  \Phi_{p}({\theta}) = -\frac{GM}{x(\theta, \omega) R_p} - \frac{1}{2} \Omega^2 (x(\theta, \omega) R_p)^2 \sin^2 \theta, \label{eq:potential}
 \end{equation}
where $x(\theta, \omega) = R_{*}(\theta, \omega)/R_{p}$,  and $R_{p}$ is the polar radius. The radial profile of the stars was determined by solving the following cubic potential equation \citep{Collins-Harrington-1966ApJ...146..152C}, which depends on $\theta$, $R_{p}$, and $\omega$ \citep{Levenhagen2024A&A...685A..57L}:


\begin{equation} R_{*}(\theta, \omega) = \frac{3R_{p}}{\omega \sin{\theta}} \cos{\left[\frac{\pi + \arccos{\omega \sin{\theta}}}{3}\right]}. \label{eq:radius} \end{equation}

Additionally the \texttt{ZPEKTR} code supports four different LD laws; in this work, we adopted the logarithmic LD law \citep{Klinglesmith1970},

\begin{equation} I\left(\mu\right) = I\left(1\right) \left[1 - \epsilon \, (1 - \mu) - w \, \mu \, \ln{\mu}\right], \label{eq:LD} \end{equation}
where $I(\mu)$ is the specific intensity,  based on the grid corresponding to the local $T_{\rm{eff}}$; $\log{g}$, $\epsilon$, and $w$ are LD coefficients taken from \cite{Klinglesmith1970}; and $\mu$ is the cosine of the angle between the line of sight and the surface normal.

Using these formulations, the \texttt{ZPEKTR} code generates prominent output parameters, including radius, effective temperature, and surface gravity ($\log g$), as functions of colatitude $\theta$. These local parameters are mapped onto a Delaunay triangular mesh over the equipotential surface of the star. The specific intensity mesh is interpolated from the non-local thermodynamic equilibrium (NLTE) TLUSTY or SYNSPEC models depending on the effective temperature value at each point in the mesh \citep{Hubeny1988CoPhC..52..103H, Hubeny1995ApJ...439..875H}.

The code outputs a normalized spectrum ranging from 4000\,\AA\ to 7000\,\AA\ with a wavelength step of 0.1\,\AA\, along with 15 stellar parameters. These are the polar, equatorial, and global mean values of the effective temperature ($T_{\rm{eff}}$), surface gravity ($\log g$), and radius ($R$), as well as mass ($M$), projected rotational velocity ($v\sin{i}$, where $v$ is the equatorial rotational velocity), rotation rate ($v/v_{c}$, where $v_{c}$ is the critical equatorial rotational velocity), angular rotation rate ($\Omega/\Omega_{c}$), inclination angle ($i$), and luminosity ($L$). 

\begin{figure}[h!] \centering \includegraphics[width=\hsize]{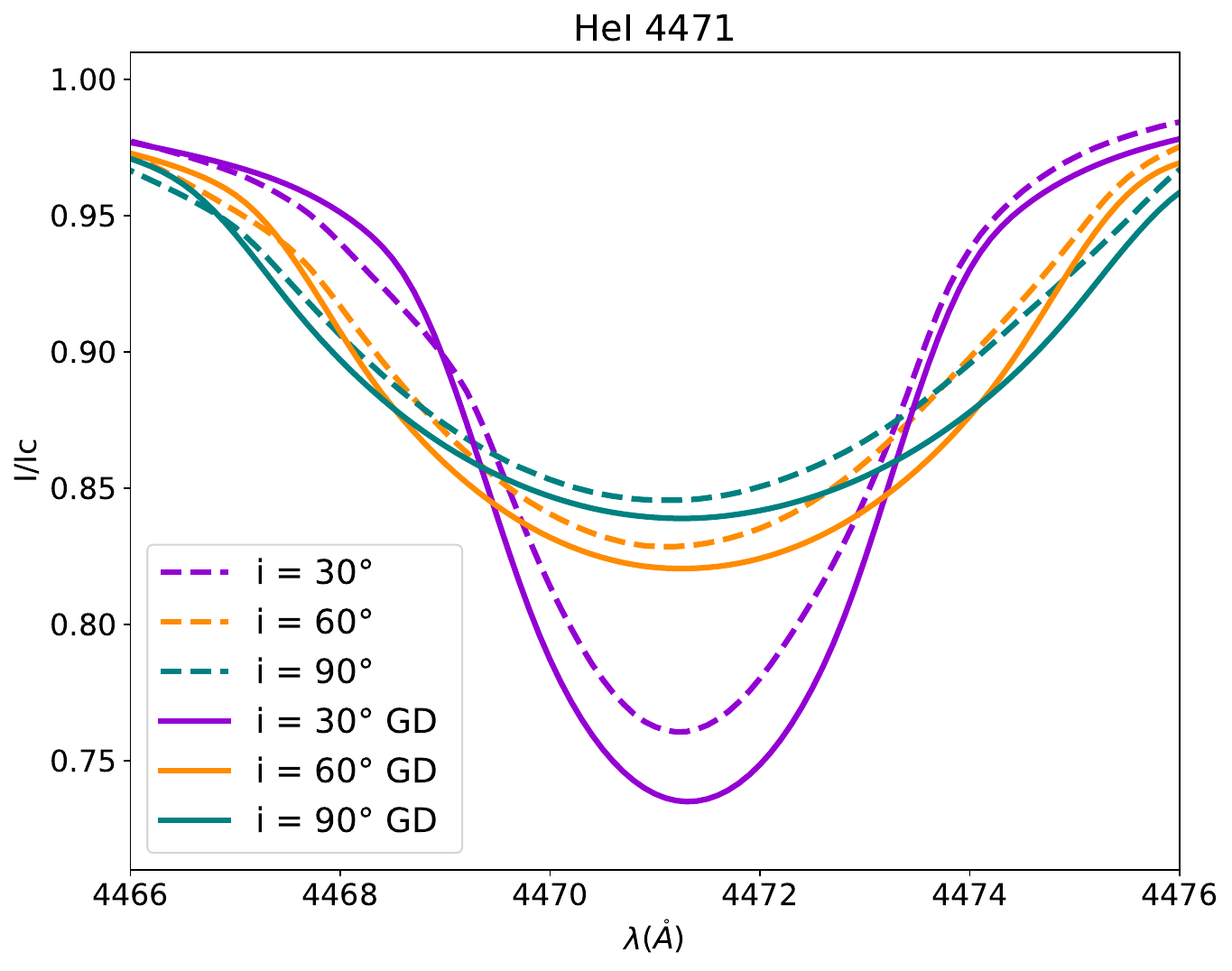} 
\caption{Comparison of the TLUSTY models (dashed lines), which do not account for stellar geometrical deformation or gravity darkening, with the \texttt{ZPEKTR} models (solid lines), which do, for three inclination angles ($30^{\circ}$, $60^{\circ}$, and $90^{\circ}$). The comparison focuses on the He I 4471 \AA\ line profile.} \label{fig:heI4471_i} \end{figure}

Fig.\ref{fig:heI4471_i} illustrates the variation in the HeI 4471 \AA\ line profile for three \texttt{ZPEKTR} and three TLUSTY models, covering inclination angles from $15^{\circ}$ to $90^{\circ}$. The input parameters for these models are $M = 10\,M_{\odot}$, $R_{\rm{pole}} = 8\,R_{\odot}$, $R_{\rm{eq}} = 9.6\,R_{\odot}$, $T_{\rm{pole}} = 20000\mathrm{K}$, and $v/v_{c} = 0.71$. The observed line profile changes reflect the influence of the varying inclination angles.

\subsection{$\chi^{2}$ test}
\label{sec:method_chi2}

To determine the best-fitting models for each star, we generated a grid of synthetic spectra using the \texttt{ZPEKTR} code. This grid was constructed based on the stellar parameters from \cite{Arcos2018}. For our sample of Be stars, assuming the logarithmic LD relation \citep{Klinglesmith1970} and when fitted to the observations, this \texttt{ZPEKTR} model enabled us to determine several stellar parameters (for more details, see \citealt{Levenhagen2014} and \citealt{Levenhagen2024A&A...685A..57L}).
We applied a  $\chi^{2}$ test to evaluate the goodness-of-fit between the synthetic spectra and the observed HeI 4471 \AA\ line profiles. The $\chi^{2}$ statistic was calculated as
\begin{equation} \chi^{2} = \sum_{j=1}^{N} \frac{(obs_{j} - model_{j})^{2}}{model_{j}},
\label{eq:chi} \end{equation}
where $obs_{j}$ indicates the normalized flux of the observed spectrum at wavelength $j$, and $model_{j}$ represents the corresponding flux of the synthetic spectrum generated by \texttt{ZPEKTR}. Both observed and synthetic fluxes were interpolated into a common wavelength grid. The summation was performed for each star over all $N$ wavelength points within a specified range, supplementing the HeI 4471 \AA\ line.

Based on their $\chi^{2}$ values, we sorted the synthetic \texttt{ZPEKTR} models in ascending order. The model that produced the lowest $\chi^{2}$ was considered the best fit to the observed HeI 4471\,\AA\, line profile. We selected the subset of models within the first quartile ($q_1$) of the $\chi^{2}$ distribution for each star to analyse the results for the ten Be stars. This approach ensured that only the models that provided the closest agreement with the observed data were considered, while also accounting for minor variations in the fitting process. We then used the selected models within $q_1$ to calculate the standard deviation of the flux at each wavelength point relative to the model with the minimum $\chi^{2}$. This standard deviation estimated the uncertainty in the derived stellar parameters.

In the particular case of the Be star HD 209409, the $\chi^{2}$ values within $q_1$ of the \texttt{ZPEKTR} models showed negligible variation. To ensure a robust error estimate of the stellar parameters for this star, we extended the range to include models up to the median ($q_2$) of the $\chi^{2}$ distribution. This adjustment allowed us to adequately account for the uncertainties in cases where the $\chi^{2}$ values were strongly clustered.

\section{Results}
\begin{figure}[h!]
    \centering
    \includegraphics[width=\hsize]{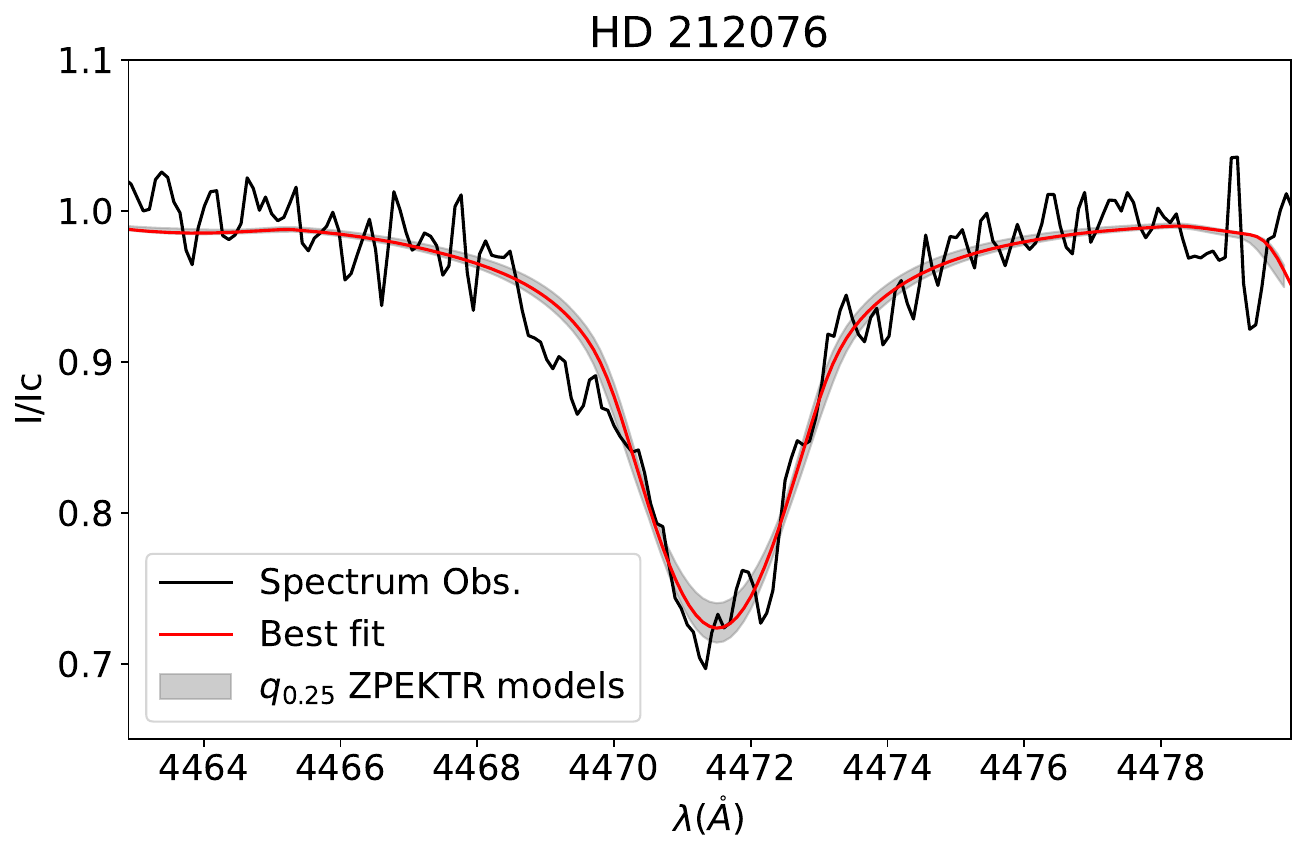}
    \caption{Observed spectrum of the photospheric line HeI 4471 \AA\ (black) for the star HD 212076. The red line shows the best-fit \texttt{ZPEKTR} model, while the grey band represents \texttt{ZPEKTR} models within the first quartile, $q_{0.25}$.}
    \label{fig:hd212076}
\end{figure}

We obtained the following results for our Be star sample using the \texttt{ZPEKTR} code and the methodology described above. To illustrate the results, the model fit for the star HD 212076 (B3III) is shown in Fig.~\ref{fig:hd212076}. We calculated the standard deviation of the flux at each wavelength point in relation to the model with the lowest $\chi^2$, and chose the models that fit the best within the first quartile of the $\chi^2$ values. The stellar parameters of this best-fitting model are summarized in Table~\ref{tab:teff_logg}: effective temperature and $\log g$ values at the pole, equator, and globally averaged throughout the stellar surface of the ten Be stars of our sample. The best-fit parameters and their associated standard deviations were determined as described in Sect.~\ref{sec:method_chi2}.

\begin{table*}
\caption{Stellar parameters for the ten Be stars obtained using the \texttt{ZPEKTR} models.}             
\label{tab:teff_logg}      
\centering        
\renewcommand{\arraystretch}{1.5}
\begin{tabular}{l|ccccccc} \hline \hline
    Star & $\langle T_{\rm{eff}} \rangle$ & $T_{\rm{eff}}^{\rm{pole}}$ & $T_{\rm{eff}}^{\rm{eq}}$ & $\langle \log g \rangle$ & $\log g_{\rm{pole}}$ & $\log g_{\rm{eq}}$  \\
    & \small[K] & \small[K] & \small[K] & \small[dex] & \small[dex] & \small[dex] \\ \hline
HD 35439 & 24605 $\pm$ 534 & 26000 $\pm$ 528 & 22455 $\pm$ 569 & 3.78 $\pm$ 0.05 & 3.84 $\pm$ 0.05 & 3.68 $\pm$ 0.05 \\
HD 41335 & 23737 $\pm$ 638 & 25000 $\pm$ 642 & 20650 $\pm$ 476 & 3.87 $\pm$ 0.08 & 3.92 $\pm$ 0.07 & 3.72 $\pm$ 0.07 \\
HD 45725 & 20786 $\pm$ 546 & 22000 $\pm$ 555 & 18080 $\pm$ 582 & 3.75 $\pm$ 0.04 & 3.81 $\pm$ 0.04 & 3.60 $\pm$ 0.05 \\
HD 60606 & 19095 $\pm$ 656 & 20000 $\pm$ 678 & 17482 $\pm$ 637 & 3.94 $\pm$ 0.07 & 3.99 $\pm$ 0.08 & 3.85 $\pm$ 0.08 \\
HD 68980 & 18104 $\pm$ 1618 & 20000 $\pm$ 1694 & 16520 $\pm$ 1306 & 3.32 $\pm$ 0.33 & 3.43 $\pm$ 0.33 & 3.23 $\pm$ 0.34 \\
HD 158427 & 18668 $\pm$ 1395 & 20000 $\pm$ 1234 & 17273 $\pm$ 1288 & 4.31 $\pm$ 0.15 & 4.38 $\pm$ 0.15 & 4.22 $\pm$ 0.14 \\
HD 209409 & 11145 $\pm$ 498 & 11500 $\pm$ 578 & 10408 $\pm$ 427 & 3.63 $\pm$ 0.13 & 3.66 $\pm$ 0.12 & 3.55 $\pm$ 0.14 \\
HD 212076 & 17374 $\pm$ 631 & 19000 $\pm$ 609 & 16608 $\pm$ 613 & 3.86 $\pm$ 0.15 & 3.96 $\pm$ 0.15 & 3.82 $\pm$ 0.15 \\
HD 212571 & 27279 $\pm$ 719 & 29000 $\pm$ 716 & 25946 $\pm$ 805 & 4.20 $\pm$ 0.03 & 4.27 $\pm$ 0.03 & 4.15 $\pm$ 0.03 \\
HD 214748 & 11917 $\pm$ 257 & 12500 $\pm$ 297 & 10737 $\pm$ 293 & 3.53 $\pm$ 0.12 & 3.58 $\pm$ 0.12 & 3.42 $\pm$ 0.11 \\ \hline
\end{tabular}
\end{table*}

Additional stellar parameters derived from the \texttt{ZPEKTR} models are presented in Table~\ref{tab:r_v_i}, complementing those listed earlier. These include the polar and equatorial radii ($R_{\rm{pole}}$ and $R_{\rm{eq}}$),  which account for the geometric deformation caused by rapid rotation, as well as the projected rotational velocity ($v\sin{i}$), equatorial and angular rotation rate ($v/v_{c}$ and $\Omega/\Omega_{c}$), and inclination angle ($i$). The final column of the table provides the inclination angles determined from interferometric observations \citep{Cochetti2019}.  

\begin{table*}
\caption{Stellar parameters for the ten Be stars obtained using the \texttt{ZPEKTR} models. The column heading $i_{\rm{Cochetti}}$ corresponds to inclination angles from \cite{Cochetti2019}.}             
\label{tab:r_v_i}      
\centering        
\renewcommand{\arraystretch}{1.5}
\begin{tabular}{l|ccccccc} \hline \hline
    Star & $R_{\rm{pole}}$ & $R_{\rm{eq}}$ & $v\sin{i}$ & $v/v_{c}$ & $\Omega/\Omega_{c}$ & $i$ & $i_{\rm{Cochetti}}$ \\
    & \small[$R_{\odot}$] & \small[$R_{\odot}$] & \small[km/s] & & & [°] & [°] \\ \hline
HD 35439 & 8.0 $\pm$ 0.6 & 9.6 $\pm$ 0.7 & 302 $\pm$ 11 & 0.71 $\pm$ 0.02 & 0.88 $\pm$ 0.02 & 58 $\pm$ 3 & 55 $\pm$ 5 \\
HD 41335 & 6.0 $\pm$ 0.5 & 7.6 $\pm$ 0.6 & 367 $\pm$ 30 & 0.79 $\pm$ 0.02 & 0.94 $\pm$ 0.01 & 75 $\pm$ 6 & 68 $\pm$ 5 \\
HD 45725 & 8.0 $\pm$ 0.4 & 10.1 $\pm$ 0.5 & 360 $\pm$ 12 & 0.79 $\pm$ 0.02 & 0.94 $\pm$ 0.01 & 68 $\pm$ 5 & 72 $\pm$ 5 \\
HD 60606 & 5.0 $\pm$ 0.6 & 5.9 $\pm$ 0.7 & 298 $\pm$ 13 & 0.68 $\pm$ 0.03 & 0.86 $\pm$ 0.02 & 66 $\pm$ 5 & 70 $\pm$ 10 \\
HD 68980 & 12.0 $\pm$ 3.8 & 15.1 $\pm$ 4.9 & 161 $\pm$ 13 & 0.79 $\pm$ 0.02 & 0.94 $\pm$ 0.01 & 32 $\pm$ 6 & 22 $\pm$ 3 \\
HD 158427 & 4.0 $\pm$ 0.6 & 4.8 $\pm$ 0.7 & 328 $\pm$ 15 & 0.71 $\pm$ 0.05 & 0.88 $\pm$ 0.04 & 44 $\pm$ 12 & 45 $\pm$ 5 \\
HD 209409 & 6.0 $\pm$ 0.8 & 6.8 $\pm$ 1.0 & 211 $\pm$ 26 & 0.60 $\pm$ 0.09 & 0.79 $\pm$ 0.08 & 82 $\pm$ 8 & 70 $\pm$ 5 \\
HD 212076 & 6.0 $\pm$ 1.4 & 7.1 $\pm$ 1.6 & 112 $\pm$ 8 & 0.68 $\pm$ 0.03 & 0.86 $\pm$ 0.03 & 19 $\pm$ 4 & 22 $\pm$ 5 \\
HD 212571 & 5.0 $\pm$ 0.1 & 5.8 $\pm$ 0.1 & 254 $\pm$ 16 & 0.63 $\pm$ 0.03 & 0.82 $\pm$ 0.02 & 38 $\pm$ 2 & 34 $\pm$ 10 \\
HD 214748 & 6.0 $\pm$ 1.2 & 7.3 $\pm$ 1.3 & 218 $\pm$ 18 & 0.72 $\pm$ 0.06 & 0.89 $\pm$ 0.05 & 69 $\pm$ 8 & 73 $\pm$ 10 \\ \hline
\end{tabular}
\end{table*}

To validate our results for the inclination angles, we compared them with interferometric measurements from \cite{Cochetti2019} using a linear regression model. In this model, the inclination angles derived from \texttt{ZPEKTR} models are treated as the independent variable ($x$), while the interferometric values serve as the dependent variable ($y$). The linear fit accounted for errors in both variables, as described by \cite{Carroll_2006} and \cite{Wasserman2007-rm}. This approach avoids the underestimation of parameters due to neglected uncertainties, as discussed by \cite{HBruzzone_1998}.

\begin{figure}[h!]
   \centering
   \includegraphics[width=\hsize]{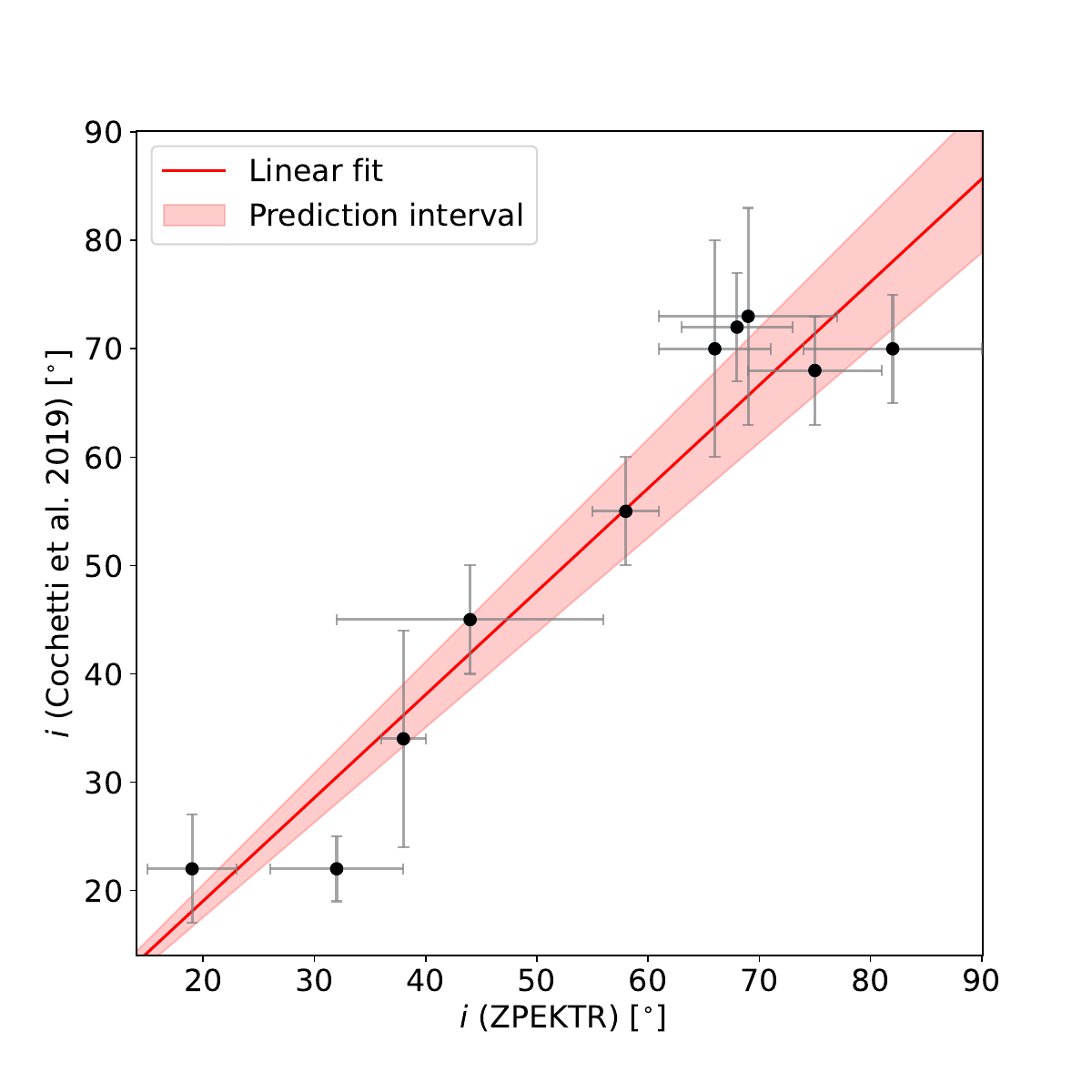}
   \caption{Comparison of inclination angles derived from \texttt{ZPEKTR} models with interferometric measurements from \cite{Cochetti2019}. The red line represents the linear fit, and the shaded region indicates the confidence interval.}
   \label{fig:inclination-angle}
\end{figure}

We fitted a linear model with no intercept ($y = m\, x $) as the intercept is not substantially different from zero. Fig.~\ref{fig:inclination-angle} shows the linear fit, with the best-fit relationship, represented by the red line, given by:
\begin{equation}
    y = 0.952\, (\pm 0.033)\, x,
    \label{eq:linear_fit}
\end{equation}
yielding a Pearson correlation coefficient $R^2 = 0.989$, indicating a strong correlation between the inclination angles derived using the present method and those obtained from previous measurements.

\section{Discussion and conclusions}
Determining inclination angles through interferometric techniques is a well-established but resource-intensive method, often restricted to a small subset of nearby and luminous stars because of spatial resolution constraints. As demonstrated in other works \citep[see e.g.][]{DomicianodeSouza2003, CheX2011, Cochetti2019}, these techniques require high-resolution observations and precise instrumentation, making them both costly and technically challenging. While they provide highly accurate results, the complexity of the setup and the extended observation times needed limit their applicability compared to alternative methods. In this work, we show that spectroscopic techniques, implemented via the \texttt{ZPEKTR} code, offer a robust and efficient alternative for estimating inclination angles in Be stars, with the added advantage of being more accessible and easier to obtain. By incorporating advanced models of gravity darkening and stellar deformation, the \texttt{ZPEKTR} code accurately reproduces observed spectral profiles, and yields results in excellent agreement with interferometric measurements, as evidenced by the strong correlation ($R^{2} = 0.989$) with the values reported by \cite{Cochetti2019}.

The slope of the linear fit, 0.952 $\pm$ 0.03, is nearly equal to unity, which demonstrates a strong concordance between the two methods. The prediction bands are also narrower towards the origin, indicating that the models are accurate for stars with smaller inclination angles \citep{motesinos2024}.

For star HD 209409, where the $\chi^{2}$ values exhibited minimal variation within the first quartile, we adjusted our methodology to include models up to the median of the $\chi^{2}$ distribution. This flexibility highlights the adaptability of the ZPEKTR framework in addressing cases where tightly clustered solutions might otherwise underestimate uncertainties.

The code effectively mitigates the challenges posed by line broadening and blending, which typically increase the uncertainty of determined $v\sin i$ values \citep{Levenhagen2014} and the angle of inclination at higher rotation rates. This accuracy stems from the ability of the code to model polar and equatorial temperature variations and account for the oblate shape of the star. These features are often overlooked in simpler spectroscopic approaches. We also assumed a logarithmic LD law for our analysis because it produced slightly better results than a linear LD law. However, the linear LD law would also yield acceptable results, which demonstrates the flexibility of the method.

Furthermore, the inclination angle  ($i)$ critically affects the determination of $v\sin i$. Small errors in $i$ estimations can lead to substantial discrepancies in $v\sin i$ for higher rotational velocities due to the magnification of projection effects. Gravity darkening and centrifugal flattening for rapidly rotating massive stars introduce additional complexities, as these phenomena influence line profiles differently across the stellar surface  \citep{VonZeipel}. We compared the projected rotational velocities derived using \texttt{ZPEKTR} with those in the BeSOS database, where plane-parallel models are assumed \citep{Arcos2018}; our results consistently show higher $v\sin i$ values, approximately 20$\%$ higher, while the effective temperatures are similar.

Furthermore, our $v\sin i$ values exceeded those of \cite{Abt2002ApJ...573..359A} by 18\%, based on Full Width Half Maximum (FWHM) calibrations using \cite{Slettebak1975ApJS...29..137S} standard stars. These findings suggest that gravity-darkening effects may systematically lead to underestimations in classical approaches.
Compared to traditional plane-parallel atmospheric models, the \texttt{ZPEKTR} code reduces dispersion in $v\sin i$ measurements compared to classical plane-parallel atmospheric models, which suggests that gravity-darkening effects may systematically lead to underestimations in classical approaches.

In general, our study highlights the advantages of \texttt{ZPEKTR} for determining the angle of inclination and the rotational velocities in Be stars.
\begin{itemize}
\item The \texttt{ZPEKTR} code offers a resource-efficient alternative to interferometry for determining the inclination angles in Be stars and provides precise measurements that are in excellent agreement with interferometric results.  This allows for more extensive surveys, and as such the code can be applied to a wider range of stellar populations, including those that are not easily observable with interferometry.
    \item The method incorporates the logarithmic LD law, which provides better accuracy than the linear LD law, although both laws are acceptable.
    \item ZPEKTR effectively reduces the uncertainty in $v\sin i$ measurements, offering consistently higher values compared to traditional methods such as those from \cite{Arcos2018} and \cite{Abt2002ApJ...573..359A}, which indicates that classical methods may systematically underestimate $v\sin i$.
\end{itemize}

In this work, we modelled the HeI 4471 \AA\, spectral line. Future work will extend this methodology to additional photospheric lines, such as HeI 4387 \AA, HeI 4713 \AA, and HeI 5015 \AA, since \cite{Solar2022} demonstrate that $v\sin i$ measurements can vary depending on the chosen spectral features. To improve our analysis, we are currently working on implementing a Bayesian neural network for the better estimation of errors, which will enhance the reliability and precision of our results.


\begin{acknowledgements}

This work used the BeSOS catalogue, operated by the IFA-UV, Chile: http://besos.ifa.uv.cl, and is maintained by Fondecyt Nº 11190945. DT acknowledges support from \emph{ANID BECAS/DOCTORADO NACIONAL/2024-21240465}. MC \& CA acknowledge partial support from Centro de Astrofísica de Valparaíso. MC, IA \& CA thank the support from FONDECYT project No.~1230131, and from ANID / FONDO 2023 ALMA / 31230039. AC acknowledges partial support from Centro de Estudios Atmosféricos y Cambio Climático, CIDI CR 212.115.013. AC thanks the support from Proyecto PUENTE, financiado por el proyecto PFE UVA20993, Fortalecimiento del Sistema de Investigación e Innovación de la Universidad de Valparaíso, Universidad de Valparaíso. This work has been possible thanks to the use of AWS-U.Chile-NLHPC credits. Powered@NLHPC: This research was partially supported by the supercomputing infrastructure of the NLHPC (ECM-02). This project has received funding from the European Union’s Framework Programme for Research and Innovation Horizon 2020 (2014-2020) under the Marie Skłodowska-Curie Grant Agreement No.~823734 and is also Co-funded by the European Union (Project 101183150 - OCEANS)

\end{acknowledgements}

\bibliographystyle{aa} 
\bibliography{reference}

\end{document}